
\magnification = \magstep1
\baselineskip = 24 true pt

\rightline{IMSC-93/11}

\rightline{SBNC-06/93}

\rightline{hepth/9306126}

\bigskip

\centerline {$SL(n,R)$ KDV HIERARCHY AND ITS NONPOLYNOMIAL REALIZATION}
\centerline {THROUGH KAC MOODY CURRENTS}
\vskip 2cm

\centerline {Sasanka Ghosh$^1$ and Samir K. Paul$^2$}
\vskip 2cm

\centerline {1. {\it Institute of Mathematical Sciences}}
\centerline {{\it C.I.T. Campus, Madras - 600 113, India}}
\bigskip

\centerline {2. {\it S.N. Bose National Centre For Basic Sciences}}
\centerline {{\it DB 17, Sector 1, Salt Lake, Calcutta - 700 064, India}}

\vskip 4cm

It is shown that $SL(n,R)$ KdV hierarchy can be expressed as definite
nonpolynimials in Kac Moody currents and their derivatives by the
action of Borel subgroup of $SL(n,R)$ on the phase space of centrally
extended $sl(n,R)$ Kac Moody currents. Construction of Lax pair is
shown, confirming Drinfeld Sokolov type Hamiltonian reduction. This
suggests an example of a moduli space with simplectic structure
corresponding to extended conformal symmetries.

\vfill\break

Drinfeld  Sokolov Hamiltonian reduction procedure [1] in the theory of
nonlinear integrable systems has opened up the most interesting aspect
{\it viz. } the connection between conformal field theory and
intergable systems [2,3,4]. This came out via the correspondence
between Poisson bracket algebra among the KdV type fields and the
quantum Dirac bracket algebra among normal ordered quantities. The
direction was from quantum algebra to Poisson bracket algebra in
$\hbar \rightarrow 0$  limit [2]. This was the case due to normal
ordering ambiguity in the modes of periodic nonlinear fields. From
these lines there were hints that there are hidden quantum symmetries
or symmetries reflected through extended conformal algebras or some
current algebra whose classical counterperts are the Gelfand Dikki
Poisson bracket algebras of KdV type nonlinear integrable systems.
Apart from the pioneering works of Zamolodchikov and others [5,6], the
quantum symmetries having hidden symmetries were shown to exist by
Polyakov [7] and Bershadsky and Ooguri [3]. Polyakov constructed
diffeomorphisms from restricted $SL(2,R)$ and $SL(3,R)$
transformations leading to Virasoro algebra and $W_3$ algebra. He
demonstrated the connection between the Wess Zumino Novikov Witten
(WZNW) action and its gravitational analogue implementing the above
observation in the $SL(2,R)$ case. The classical underplay in the
restricted gauge fixing procedure is essentially Drinfeld Sokolov
Hamiltonian reduction for $SL(n,R)$ which essentially rests on the
important physical principle of gauge equivalence between Lax
operators. Bradshadsky and Ooguri [3] observed that the physical
Hilbert space of the right moving sector of constrainted WZNW model
can give rise to irreducible representations of $W_n$ algebra [6].
This symmetry is again the quantum analogue of Drinfeld Sokolov
Hamiltonian reduction [1].

At this point it is worthwhile to enquire whether there are new systems
admitting Drinfeld Sokolov type Hamiltonian reduction and what kind of
simplectic structure the reduced phase space can have and to what
quantum symmetry this structure may correspond.
In this letter we want to look into a moduli space (explained below)
which is inequivalent to the usual Drinfeld Sokolov moduli space [1]
and to show that it is, in fact, classical analogue of the same
quantum symmetries, shown in the references 2 and 3.

Drinfeld Sokolov Hamiltonian reduction has been studied for
the action of the subgroup of $SL(n,R)$ of upper triangular matrices with
1's in the main diagonal (group of such matrices are denoted by N) on
the space of centrally extended $sl(n,R)$ Kac Moody currents and the
possible quantum generalization of the procedure [2,3]. The emerging
$sl(n,R)$ classical KdV
fields in this case are polynomials in currents and their derivatives.

In this letter we enlarge the action of the above subgroup by taking
Borel subgroup ${\tilde N}$ of $SL(n,R)$\footnote\dag{Borel subgroup
$SL(n,R)$is the group generated by $\{H_i, J_l^+; \quad
i=1,2,..,n-1; l=1,2,..,,{1\over 2}n(n-1)\}$ where $\{H_i,
J_l^{\pm}\}$is the standard Cartan Weyl basis of $sl(n,R)$ algebra. If
we denote this group by $\tilde N$, $N$, the group of upper traingular
matrices of $SL(n,R)$ with 1's in the main diagonal and generated
by$\{J_l^+\}$ is a subgroup of $\tilde N$: $N \subset {\tilde N}
\subset SL(n,R)$. Sometimes $N$ is also termed as Borel subgroup in
the literature.} and show that $sl(n,R)$ KdV
fields are definite {\it rational functions } (not polynomials) of the
Kac Moody currents and their derivatives. Subsequently our
construction of canonical Lax equation, which is gauge equivalence
preserving [1], ensures Hamiltonian reduction. We also discuss the
corresponding quantum symmetry of the classical KdV system. We
explicitly demonstrate our formulation for $SL(2,R)$ and $SL(3,R)$.

Consider the space of first order differential operators

$$ {\cal L} = k {d\over {dx}} + v(x), \eqno (1) $$

\noindent $v(x)$ taking values in $sl(n,R)$. ${\cal L}$ is said to be
equivalent to $S^{-1} {\cal L} S$; $S \in C^{\infty }(S^1, SL(n,R))$,
which implies

$$ v(x) \sim  S^{-1} v(x) S + k S^{-1} \partial_x S . \eqno (2) $$

\noindent In particular, if $S \in C^{\infty }(S^1, \tilde N)$,
$\tilde N$ being
the Borel subgroup of $SL(n,R)$, there is a unique $S$ such that

$$S^{-1} v(x) S + k S^{-1} \partial_x S = \sum ^n _{i=1} w_{n+1-i} e_{in} +
\Lambda , \eqno (3) $$

\noindent where

$$ \Lambda = \sum ^{n-1}_{i=1} e_{i+1,i} \quad ; \quad w_{1} = 0  \eqno (4) $$

\noindent and $e_{ij}$ denotes the $n \times n$ matrix with 1 in $(i,j)$ {\it
th.}
position and zero elsewhere. $w$'s in (3) are rational functions in
the elements of $v(x)$ and their derivatives. $w_n, w_{n-1},....,w_2$
are gauge invariant quantites {\it i.e.} ${\delta}_S w_i = 0$, \quad
\quad $i = 2,3,...,n$ for any $S \in C^{\infty}(S^1, \tilde N)$.
Relation (2) defines an action of $\tilde N$ on $\tilde M$ where
$\tilde M$ is the phase space with coordinates $\{ h_i(x) ,
j_i^{\pm}(x) , k \}$. The phase space coordinates saitsfy the centrally
extended Kac Moody algebra with central extension
$k \in R$. $v(x)$ in (1) has the form

$$v(x) = \sum^{n-1}_{i=1} h_i(x) H_i + \sum^{{1\over 2}n(n-1)}_{i=1}
j_i^+(x) J_i^+ + \sum^{{1\over 2}n(n-1)}_{i=1} j_i^-(x) J_i^-. \eqno
(5a) $$

\noindent Here $\{ H_i , J_i^{\pm }\}$ is the Cartan Weyl basis of
$sl(n,R)$.

(5a) together with (3) and (4) gives

$$ j_i^-(x) = 0 ; \quad \quad n\leq i \leq {1\over 2}n(n - 1) \eqno
(5b) $$

It is easy to verify from (2) and (5) that $w_i$'s in (3) together
with central extension $k$ are the coordinates of the moduli space
${{\tilde M} \over {\tilde N}}$. Moreover, any gauge invariant
quantity can be expressed in terms of $w_i$'s and their derivatives
only.

In order to ensure now that our procedure corresponds to Hamiltonian
reduction {\it a la'} Drinfeld Sokolov, we have to formulate the
construction of gauge equivalence preserving canonical Lax equation
[1]. For this we first assert that if

$$\tilde {\cal L}(x,\lambda) = k {d\over {dx}} + \tilde {\Lambda }(x)
+ \sum^{n-1}_{i=1} h_i(x) H_i + \sum^{{1\over 2}n(n-1)}_{i=1} j^+_i(x)
J_i^+ \eqno (6a) $$

\noindent with

$$\tilde {\Lambda }(x) = {\lambda \over {j_1^-j_2^-..j_{n-1}^-}}
e_{1n} + \sum_{i=1}^{n-1} j_i^-(x)e_{i+1,i}, \eqno (6b) $$

\noindent $\lambda $ being the spectral parameter, there is a unique

$$ T = \sum_{i=0}^{\infty } T_i(x) \lambda ^{-i} \eqno (6c) $$

\noindent with the first column of $T_0 = (1,0,...,o)^T$ and the first
column of
$T_i = (0,0,..,0)^T$  $(i \neq 0)$ such that

$$\tilde {\cal L}_0(x,\lambda) = T \tilde {\cal L}(x,\lambda) T^{-1}
\eqno (7a) $$

\noindent is of the form

$$\tilde {\cal L}_0(x,\lambda) = k {d\over {dx}} + \tilde {\Lambda}(x)
+ \sum_{i=0}^{\infty} f_i(x) \tilde {\Lambda }(x) + {k\over 2}
\sum_{i=1}^{n-1} A_i(x) H_i \eqno (7b) $$

\noindent where $f_i(x)$ are functions of $x$ and $A_i(x) = (j_i^-(x))^{-1}
\partial j_i^-(x)$ ($i=1,2,..,n-1$ and so sum over $i$). $f_i(x)$ and
$T$  can be
uniquely determined from the recurrence relations obtained from (6)
and (7). ${\tilde \Lambda }(x)$ in (6b) has the property that for
$SL(n,R)$ group

$$({\tilde \Lambda }(x))^n = {\bf E}, \eqno (7c) $$

\noindent $n \times n$ identity matrix.

Notice that $\tilde {\cal L}$ in (6) reduces to the initial linear
differential operator ${\cal L}$ (1,5) for $\lambda = 0$.
Whereas, in order to get $SL(n,R)$ KdV hierarchy one has to suitably
modify ${\cal L}$ by injecting the spectral parameter $\lambda $ as in
(6).

The Lax equation, given by

$${{d\tilde {\cal L}}\over {dt}} = [ \tilde {\cal A} , \tilde {\cal L} ]
\eqno (8) $$

\noindent will be gauge equivalence preserving if $\tilde {\cal A}$ is
chosen in such a way that both sides of (8) are independent of the
spectral prameter $\lambda $ and the time evolution of $w_i$'s from
(8) is expressed as polynomials in $w_i$'s and their derivatives only.
Following a procedure similar to that of Drinfeld Sokolov we can
choose $\tilde {\cal A}$ as

$$ \tilde {\cal A} = \sum_{i=1}^{m} C_i \Phi(\tilde {\Lambda }^i(x))^+,
\eqno (9a) $$

\noindent $m$ being positive integer and $C_i = 0$ modulo $n$ because
of (7c).
$\Phi(\tilde {\Lambda}^i(x))^+$ is the polynomial part of $\Phi(\tilde
{\Lambda}^i(x))$, where $\Phi(\tilde {\Lambda}^i(x))$ can be defined by

$$\Phi(\tilde {\Lambda}^i(x)) = T^{-1} (\tilde {\Lambda}^i(x)) T =
i \sum_{l=-\infty}^r \phi_l(x) \tilde {\Lambda}^l(x), \eqno (9b) $$

\noindent $r$ being a positive integer and $\phi_l(x)$ diagonal
matrices. $T$ in (9b) is defined in (6c). Using (6b,c) one can
determine $\phi_l(x)$ and thus can also obtain $\tilde {\cal A}$.

It is interesting to note that our procedure reduces to that of Drinfeld
Sokolov if we further impose the constraints

$$ j_1^- = j_2^- = ......= j_{n-1}^- = 1 $$

\noindent on the phase space coordinates.

We now demonstrate our construction with the examples of $SL(2,R)$ and
$SL(3,R)$.

\noindent $SL(2,R)$ case :

The first order differential operator in this case is given by

$${\cal L} = k {d\over {dx}} + \pmatrix{ h(x) & j^+(x) \cr  j^-(x) &
-h(x) \cr }  \eqno (10) $$

\noindent and the Borel subgroup $\tilde N$ is the group of matrices
$\pmatrix{ a & \alpha \cr 0 & a^{-1} \cr }$ with

$$ S = \pmatrix{ a(x) & \alpha(x) \cr 0 & a^{-1}(x) \cr } \eqno (11) $$

Substituting (10) and (11) in (3) we can solve for $a(x)$ and
$\alpha(x)$ from the relation

$$ S^{-1} \pmatrix{ h(x) & j^+(x) \cr j^-(x) & -h(x) \cr } S + S^{-1}
\partial_x S = \pmatrix{ 0 & w_2(x) \cr 1 & 0 \cr } $$

\noindent whence

$$ w_2(x) = h^2(x) + j^+(x) j^-(x) + k \partial h(x) - k h(x) A(x) +
{k^2\over 4} A^2(x) - {k^2\over 2} \partial A(x) \eqno (12) $$

\noindent with $A(x) = (j^-(x))^{-1} \partial j^-(x)$.

Notice that $w_2(x)$ is not polynomial in $h(x), j^{\pm}(x)$ unlike
the previous cases [2,4]. This suggests that upon quantization our
system would correspond to some constrained gauged WZNW model which will be
different from that considered in [3]. One can easily check that
$\delta _S w_2(x)
= 0$ for arbitrary infinitesimal $a(x)$ and $\alpha(x)$ confirming
gauge invariance of $w_2(x)$. The coordinates, $\{h(x), j^{\pm}(x), k\}$ of
$\tilde M$ satisfy $sl(2,R)$ Kac Moody algebra which induces the
Poisson bracket algebra of the coordinates $\{w_2(x), k\}$ of the
reduced phase space ${{\tilde M}\over {\tilde N}}$. Writing

$$U(x) = -{1\over k} w_2(x) \eqno (13) $$

\noindent we have

$$\{U(x) , U(y)\} = 2U(x) \partial_x \delta(x-y) + \partial_xU(x)
\delta(x-y) + {k\over 2} \partial_x^3\delta(x-y) \eqno (14) $$

\noindent which looks like Gelfand Dikki Poisson bracket of second
kind for KdV fields. We will, however, show shortly that $U(x)$,
indeed,
satisfies KdV equation. Notice that ${{\tilde M}\over N}$ is a
subspace of ${{\tilde M}\over {\tilde N}}$ since $w_2(x)$ in (12)
reduces to the same expression of the gauge invariant quantity
obtained in [2,4]
only when $j^-(x) = 1$. From (8) and (9) we have

\item {(i)} When $C_1 \neq 0$ and other $C_i$'s are zero

$$\eqalign{\partial_t h(x) &= -k\partial_x(h - {k\over 2} A) - j^+ J^-
\cr \partial_t j^+(x) &= 2j^+ (h - {k\over 2} A) \cr \partial_t j^-(x) &=
0} $$

These equations lead to

$$\partial_t U(x) = \partial_x U(x) \eqno (15a) $$

\noindent $U(x)$ being defined in (11) and (12).

\item {(ii)} When $C_3 \neq 0$ and other $C_i$'s are zero

$$\eqalign{\partial_t h(x) &= {k^3\over 4} \partial^2_xU - {k^2\over
2} \partial_xU (h - {k\over 2} A) - {k^2\over 2} U \partial_x(h -
{k\over 2} A) - {k\over 2} U j^+ j^- \cr
\partial_t j^+(x) &= - {k^2\over 2} j^+ \partial_xU + k j^+ U (h -
{k\over 2} A) \cr
\partial_t j^- &= 0} $$

These equations lead to

$$\partial_t U(x) = \partial_x^3 U(x) + 6 U(x) \partial_x U(x) \eqno
(15b) $$

\noindent after proper rescaling of $U(x)$ and $x$. (15) is the wellknown
KdV equation. Similarly for other nonzero
$C_i$'s one can obtain KdV hierarchy. It is rather important to
observe that for each $C_i$, as in (14) and (15) for $C_1$ and $C_3$,
the time evolution of the coordinate $U(x)$ (or $w_2(x)$) of ${{\tilde
M}\over {\tilde N}}$ is gauge invariant. This clearly shows a
consistent formulation of Hamiltonian reduction in our case.

\noindent $SL(3,R)$ case :

In this case the linear differential operator has the form

$${\cal L} = k {d\over {dx}} + \pmatrix{h_1(x) & j_1^+(x) & j_3^+(x)
\cr  j_1^-(x) & - h_1(x) + h_2(x) & j_2^+(x) \cr  j_3^-(x) & j_2^-(x) &
- h_2(x) } \eqno (16) $$

\noindent where $h_1(x)$, $h_2(x)$, $j_1^{\pm}(x)$, $j_2^{\pm}(x)$ and
$j_3^{\pm}(x)$ satisfy $sl(3,R)$ Kac Moody algebra. The Borel subgroup
$\tilde N$  of $SL(3,R)$ is the group of matrices
$\pmatrix{a & p & q \cr 0 & a^{-1}b & n \cr 0 & 0 & b^{-1}}$ with

$$ S = \pmatrix{a(x) & p(x) & q(x) \cr 0 & a^{-1}(x)b(x) & n(x) \cr 0
& 0 & b^{-1}(x) \cr} \eqno (17) $$

Substituting (16) and (17) in (3) we can solve for $a(x)$, $p(x)$,
$q(x)$, $n(x)$ and $b(x)$ from the relation

$$S^{-1} \pmatrix{h_1(x) & j_1^+(x) & j_3^+(x) \cr  j_1^-(x) & -
h_1(x) + h_2(x) & j_2^+(x) \cr  j_3^-(x) & j_2^-(x) & - h_2(x) \cr} S +
S^{-1} \partial_x S = \pmatrix{0 & 0 & w_3(x) \cr 1 & 0 & w_2(x) \cr 0
& 1 & 0 \cr} $$

\noindent where

$$ w_2(x) = j_1^-(x) j_2^-(x) + j_2^+(x) j_2^-(x) + {\tilde h}^2_1(x)
+ {\tilde h}_2^2(x) - {\tilde h}_1(x) {\tilde h}_2(x) + k \partial_x
{\tilde h}_1(x) + k \partial_x {\tilde h}_2(x), \eqno (18a) $$

$$\eqalign{w_3(x) & = j_1^-(x) j_2^-(x) j_3^+(x) + j_1^-(x) j_1^+(x) {\tilde
h}_2(x) - j_2^-(x) j_2^+(x) {\tilde h}_1(x) + {\tilde h}_1^2(x)
{\tilde h}_2(x) \cr & - {\tilde h}_1(x) {\tilde h}_2^2(x) + k \partial_x
(j_1^-(x) j_1^+(x)) + 2k {\tilde h}_1(x) \partial_x {\tilde h}_1(x) -
k {\tilde h}_1(x) \partial_x {\tilde h}_2(x) + k^2 \partial_x^2 {\tilde
h}_1(x)} \eqno (18b) $$

\noindent and

$$j_3^-(x) = 0 \eqno (18c) $$

\noindent with

$${\tilde h}_1(x) = h_1(x) - {2\over 3} (j_1^-(x))^{-1} \partial_x
j_1^-(x) - {1\over 3} (j_2^-(x))^{-1} \partial_x j_2^-(x) $$

\noindent and

$${\tilde h}_2(x) = h_2(x) - {1\over 3} (j_1^-(x))^{-1} \partial_x
j_1^-(x) - {2\over 3} (j_2^-(x))^{-1} \partial_x j_2^-(x) $$

Unlike the previous cases [2,4] again $w_2(x)$ and $w_3(x)$ are
non polynomials in  Kac Moody currents. One can easily check that

$$\eqalign{\delta_S w_2(x) & = 0 \cr \delta_S w_3(x) & =0} \eqno (19)
$$

\noindent for arbitrary infinitesimal $a(x)$, $b(x)$, $p(x)$, $q(x)$
and $n(x)$, confirming the gauge invariance of $w_2(x)$ and $w_3(x)$.
It can also be shown that $sl(3,R)$ Kac Moody algebra among the
coordinates, $h_1(x)$, $h_2(x)$, $j_1^{\pm}(x)$, $j_2^{\pm}(x)$
$j_3^{\pm}(x)$ and $k$ induces the Poisson brackets of the
coordinates, $w_2(x)$, $w_3(x)$ and $k$ of the phase space ${{\tilde
M}\over {\tilde N}}$.

\noindent Writing

$$\eqalign{W_2(x) & = {1\over k} w_2(x) \cr W_3(x) & = {1\over k} w_3(x)}
 \eqno (20) $$

\noindent we have

$$\{ W_2(x) , W_2(y) \} = [W_2(x) + W_2(y)] \partial_x \delta(x-y) + 2k
\partial_x^3 \delta(x-y) \eqno (21a)$$

$$\{ W_2(x) , W_3(y) \} = [W_3(x) + 2 W_3(y)] \partial_x(x-y) \eqno (21b)$$

$$\eqalign{\{ W_3(x) , W_3(y) \} = & - {1\over 6} k^3 \delta(x-y) - {1\over 3}
 [W_2^2(x)
+ W_2^2(y)] \partial_x \delta(x-y) \cr  & + {1\over 4} [\partial_x^2 W_2(x) +
\partial_x^2(y)] \partial_x \delta(x-y) - {5\over 12}k [W_2(x) + W_2(y)]
\partial_x^3 \delta(x-y)} \eqno (21c)$$

\noindent These look like the Gelfand Dikki Poisson brackets of second kind
for KdV and W fields.

$SL(3,R)$ KdV hierarchy can be obtained now from the spectral
parameterful Lax operator (6a)

$${\tilde {\cal L}}(x,\lambda) = k {d\over {dx}} + {\tilde \Lambda}(x) +
j_1^+(x) J_1^+ + j_2^+(x) J_2^+ + j_3^+(x) J_3^+ + h_1(x) H_1 + h_2(x)
H_2 \eqno (22a) $$

\noindent with

$${\tilde \Lambda}(x) = \pmatrix{0 & 0 & {\lambda \over {j_1^- j_2^-}}
\cr j_1^- & 0 & 0 \cr 0 & j_2^- & 0 \cr} \eqno (22b) $$

\noindent and from the operator in (7b),

$${\tilde {\cal L}}_0(x,\lambda) = k {d\over {dx}} + {\tilde \Lambda}(x) +
\sum^{\infty}_{i=0} f_i(x) {\tilde \Lambda}(x) + {k\over 2} A_1 H_1 +
{k\over 2} A_2 H_2 \eqno (23) $$

\noindent satisfying the relation (7a) where,
$A_1 = (j_1^-(x))^{-1} \partial_x j_1^-(x)$ and $A_2 = (j_2^-(x))^{-1}
\partial_x j_2^-(x)$. $J_1^{\pm}, J_2^{\pm}, J_3^{\pm}, H_1$ and $H_2$
in the above equations are the generators of $SL(3,R)$ group.

The coefficients $C_i$ of the series (9a) in this case is
zero modulo 3, {\it i.e.} $C_0 = C_3 = C_6 = .... = 0$. It can be shown by a
lengthy but straightforward calculation that $w_2(x)$ and
$w_3(x)$ satisfy the equations of motions when $C_4, C_5 \neq 0$ and other
$C_i$'s are zero, which can be recast into the Boussinesq equations.
Boussinesq hierarchy can be obtained in a similar way by
choosing other higher $C_i$'s to be non zero. This, in fact, confirms
the Hamiltonian reduction in $SL(3,R)$ case. We should also mention
that the expressions for $w_2(x)$ and $w_3(x)$ reduce to the similar
expressions in [4] only when $j_1^-(x) = j_2^-(x) = 1$, implying once
again ${{\tilde M}\over N} \subset {{\tilde M}\over {\tilde N}}$.

We now come to the question of quantum algebra reflecting the symmetry of a
quantum system so that the classical limit will be the Poisson bracket algebra
in the reduced phase space ${{\tilde M}\over {\tilde N}}$. Since $\omega_i$'s
$(i=2,3,...,n)$ in our case are rational functions of currents in (5) we have
to choose a siutable representation to make the rational functions into
polynomial in terms of the new fields. This can be done first by considering
the Wakimoto representation [3, 8] of Kac Moody currents and then by taking the
exponentional form of $(\beta, \gamma)$ representation, introduced by Gerasimov
{\it et. al.} [9]. For convenience, however, we develop the quantum formalism
first with the example of $sl(2,R)$ case.

The classical Wakimoto representation for $sl(2,R)$ phase space is given by

$$h(x) = \beta(x) \gamma(x) + {\sqrt {k\over 2}} \partial\phi(x) \eqno
(24a) $$

$$j^+(x) = - \beta(x) (\gamma(x))^2 - {\sqrt 2k} \gamma(x) \partial\phi(x)
- k \partial\gamma(x) \eqno (24b)$$

\noindent and

$$j^-(x) = \beta(x) \eqno (24c)$$

If we write $\beta(x)$ and $\gamma(x)$ as

$$\beta(x) = exp( - u(x) -i v(x)) \eqno (25a)$$

$$\gamma(x) = i \partial v(x) exp(u(x) + i v(x)) \eqno (25b)$$

\noindent (24) becomes

$$h(x) = i \partial v(x) + {\sqrt {k\over 2}} \partial\phi(x) \eqno (26a)$$

$$j^+(x) = [(k + 1)(\partial v(x))^2 - ik \partial^2v(x) - i{\sqrt(2k)}
\partial v(x) \partial\phi(x) - ik \partial u(x) \partial v(x)]
exp(u(x) + i v(x)) \eqno (26b)$$

\noindent and

$$j^-(x) = exp( - u(x) - i v(x)) \eqno (26c)$$

\noindent and the phase space structure is now given by

$$\eqalign{\{ \phi(x) , \partial\phi(y) \} & = \delta(x - y) \cr
\{ \beta(x) , \gamma(y) \} & = \delta(x - y) \cr
\{ u(x) , \partial v(y) \} & = i \delta(x - y)} \eqno (27)$$

(24), (26) together with (27) satisfy the classical $sl(2,R)$ Kac Moody algebra
with central extension $k$. This is, in fact, $\hbar \rightarrow 0$ limit of
the $sl(2,R)$ quantum Kac Moody algebra. Following this limiting procedure to
come from $sl(2,R)$ quantum Kac Moody algebra to $sl(2,R)$ classical Kac Moody
algebra we have definite transition  from classical to quantum fields
properly scaled by $\hbar$. Together with the limits taken in the reference 2
we have the following the transitions in (26).

$$ \eqalign{ k_{quantum} & \rightarrow - \hbar^{-1} k_{cl} \cr \phi_{quantum} &
 \rightarrow {\sqrt\hbar} \hbar^{-1} \phi_{cl} \cr u_{quantum} & \rightarrow
u_{cl} \cr v_{quantum} & \rightarrow v_{cl}} \eqno (28)$$

\noindent as $\hbar \rightarrow 0$. Using (26) the expression for $U(x)$
(12, 13) in terms of $u(x)$, $v(x)$, $\phi(x)$ takes the form

$$U(x) = - {1\over 2} [\partial\phi(x) + {\sqrt{k\over 2}} \partial (u(x) + i
v(x))]^2 - {\sqrt{k\over 2}} \partial^2[\phi(x) + {\sqrt{2\over k}} (u(x) + i
v(x))] \eqno (29)$$

This is obviously the classical limit of free field represention of the stress
tensor, $T(z)$ with back ground. Using the limits in (28) and $T(z)
\rightarrow \hbar U(x)$ as $\hbar \rightarrow 0$, we have the following form
for $T(z)$ in the quantum case.

$$T(z) = - {1\over 2} [\partial\phi(x) - i {\sqrt{{k+2}\over 2}} \partial
(u(x) + i
v(x))]^2 + {i\over {\sqrt2}} {{k+1}\over {\sqrt{k+2}}} \partial^2[\phi(x) - i
{\sqrt{{k+2}\over 2}} (u(x) + i v(x))] \eqno (30) $$

\noindent giving the central charge of the quantum theory as

$$c = {3k\over {k+2}} - 6k - 2 \eqno (31)$$

Thus the bracket algebra of the reduced phase space, ${{\tilde M}\over
{\tilde N}}$, in this case,  corresponds to the Virasoro algebra for a theory
which is unitary. Notice that the exact form of $T(z)$ in (30) is unique
subject to the condition that its classical limit is (12) and $T(z)$
corresponds to unitary, irreducible representation of Virasoro algebra.

Next we point out the procedure for $sl(3,R)$ case. In this case one can
follow exactly the same procedure as we did in $sl(2,R)$ case by taking the
explicit $(\beta, \gamma)$ representation for quantum $sl(3,R)$ Kac Moody
algebra [3] and then substituting the exponential forms for ghosts as in [9].
It is worth mentioning here that the advantage of choosing exponential
representations for ghosts becomes much more transparent in $sl(3,R)$ case.
In this representation the constraint $j^-_3(x) = 0$ in (18c) gets linearized.
We omit here the calculation since it is lengthy but straightforward. It is,
however, important to mention that the expression for $w_2(x)$ and $w_3(x)$ in
(18) are respectively the classical limit of $T(z)$ and $W_3(z)$ which
correspond to a unitary theory with a central charge for $W_3$ algebra [5].
Clearly one can proceed in the same way for $sl(4,R)$ and onwards. In each
case we have a larger moduli space ${{\tilde M}\over {\tilde N}}$ leading
to same definite quantum symmetry as in [6]. So we conclude that for Drinfeld
Sokolov procedure the maximum action is that of Borel subgroup on the Kac
Moody phase space giving rise to a moduli space whose symmetry correspond
to that of $Z_n$ [6] symmetry in the quantum theory.

Authors are thankful to P. Majumdar for discussions. One of the authors (SKP)
is thankful to P. Mitra and D. Gangopadhyay for discussions and also
acknowledges his discussions with Ryu Sasaki. SKP would like to thank
Institute of Mathematical Sciences, Madras for providing him with a visiting
position while the work was initiated and SG would like to thank S.N. Bose
National Centre for Basic Sciences, Calcutta for providing him with a visiting
 position
to continue the work.

\vfill\break

\noindent References :

\item {1.} V.G. Drinfeld and V.V. Sokolov; Jour. Sov. Math. {\bf 30} (1984)

1975.

\item {2.} A. Belavin; KdV type equations and W- algebras, Hand written

manuscript, 1988.

\item {3.} M. Bershadsky and H. Ooguri, Comm. Math. Phys. {\bf 126} (1989) 49.

\item {4.} I. Bakas, Phys. Lett. {\bf B219} (1989) 283.

\item {5.} A.B. Zamolodchikov and V.A. Fateev, Nucl. Phys. {\bf B280 [FS 18]}

(1987) 644.

\item {6.} V.A. Fateev and S.K. Lykyanov, Int. Jour. Mod. Phys. {\bf A3}

(1988) 507.

\item {7.} A.M. Polyakov, Int. Jour. Mod Phys {\bf A5} (1990) 833.

\item {8.} M. Wakimoto, Comm. Math. Phys. {\bf 104} (1986) 604; V.I.S.

Dotsenko, Nucl. Phys. {\bf B338} (1990) 747.

\item {9.} A. Gerasimov, A. Marshakov and A. Morozov; preprint ITEP 89-139.

\vfill\eject\end